\documentclass[letterpaper]{article} 
\usepackage{aaai24}  
\usepackage{times}  
\usepackage{helvet}  
\usepackage{courier}  
\usepackage[hyphens]{url}  
\usepackage{graphicx} 
\usepackage{natbib}  
\usepackage{caption} 
\usepackage{algorithm}
\usepackage{algorithmic}

\usepackage{newfloat}
\usepackage{listings}
\DeclareCaptionStyle{ruled}{labelfont=normalfont,labelsep=colon,strut=off} 
\floatstyle{ruled}
\newfloat{listing}{tb}{lst}{}
\floatname{listing}{Listing}
\usepackage{amsfonts}       
\usepackage{nicefrac}       
\usepackage{microtype}      
\usepackage{xcolor}         
\usepackage[rightcaption]{sidecap}
\usepackage{amsmath}
\usepackage{amssymb}
\usepackage{mathtools}
\usepackage{amsthm}

\title{Randomized algorithms for precise measurement of differentially-private, personalized recommendations}

\author{%
  Allegra Laro, Yanqing Chen, Hao He, Babak Aghazadeh \\
  Apple\\
  \texttt{\{allegra\_latimer, yanqing\_chen2, hao\_he2, baghazadeh\}@apple.com} 
   \\
}

\begin{document}

\maketitle

\begin{abstract}
  Personalized recommendations form an important part of today’s internet ecosystem, helping artists and creators reach interested users, and helping users discover new and engaging content. However, many users today are skeptical of platforms that personalize recommendations, in part due to historically careless treatment of personal data and data privacy. Now, businesses that rely on personalized recommendations are entering a new paradigm, where many of their systems must be overhauled to be privacy-first. In this article, we propose an algorithm for personalized recommendations that facilitates both precise and differentially-private measurement. We consider the advertising setting and conduct offline experiments to quantify how the proposed privacy-preserving algorithm affects key metrics related to user and creator experience and platform revenue compared to extremes of (private) non-personalized or non-private, personalized implementations. 
\end{abstract}

\section{Introduction}
\label{introduction}

Personalized recommendations form an important part of today’s internet ecosystem, helping artists and creators reach interested users, and helping users discover new content. Allowing the specific tastes of users to influence the content they are shown can improve the experience of all parties: creators can efficiently reach more interested users, and users receive a curated experience. However, personalization has often come at the price of user privacy. To know which recommendations users are most likely to engage with, sophisticated digital tracking infrastructures have emerged, whereby large amounts of personal information are stored and processed, often without users' explicit knowledge or consent \citep{toubiana2010adnostic}. Recently, regulations, such as GDPR and CCPA, have been passed to protect users, requiring more stringent treatment of privacy by businesses. 

Highly relevant, personalized recommendations should not have to come at the cost of privacy. Already, many tools are being developed to build private predictive models and recommender systems using data that stays on users' devices \citep{mcmahan2017communication, abadi2016deep, mcsherry2009differentially}. However, these approaches generally assume that the recommendation phase takes place privately, i.e., not being revealed to the server. While private federated measurement can be used to measure the statistics of recommended content in aggregate \citep{mcmillan2022private, erlingsson2014rappor}, these measured statistics will necessarily have some noise added to them. Unfortunately, in many cases, precisely measured statistics are critical to the business, rendering these noisy measurement approaches infeasible. For example, an on-demand content (e.g., music, television, video, or movies) application must know how many times a piece of content was played to be able to pay the content creator appropriately. Similarly, a digital advertising platform must know how many times an ad was shown to be able to charge the advertiser. The addition of noise to the number of transacted goods and resultant payments in these settings is not legally permissible. 

In this work, we posit that there need not be a trade-off between the measurement precision of personalized recommendations and the protection of the private data used in the personalized recommendation process (though the trade-off will remain for any downstream engagement metrics). In particular, we propose a differentially-private approach to measuring personalized recommendations that protects this data by adding noise in the recommendation process itself, i.e., in deciding which piece of content to show, rather than subsequently in the measurement process. By pulling the addition of noise forward in the process, from measurement into content selection, the platform is able to pay (charge) content creators (advertisers), maintain records of payment events, and provide statistics to creators (advertisers), even those with relatively low traffic, all with exact precision and strong privacy guarantees. Generally, the method we present allows for precise measurement of the outputs of on-device models whose inputs are private user features, while maintaining formal differential privacy guarantees that the private user features not be revealed to the server.The inclusion of noise in the recommendation process also yields ``exploration'' benefits, which are particularly useful in partial feedback settings like recommender systems. Non-deterministic selection rules facilitate more advanced counterfactual simulation using log data, improvements in model training by reducing selection bias, and possibly even improvements in long-term user experience by surfacing more diverse recommendations.

\begin{figure*}[t]
\centering
\includegraphics[width=\textwidth]{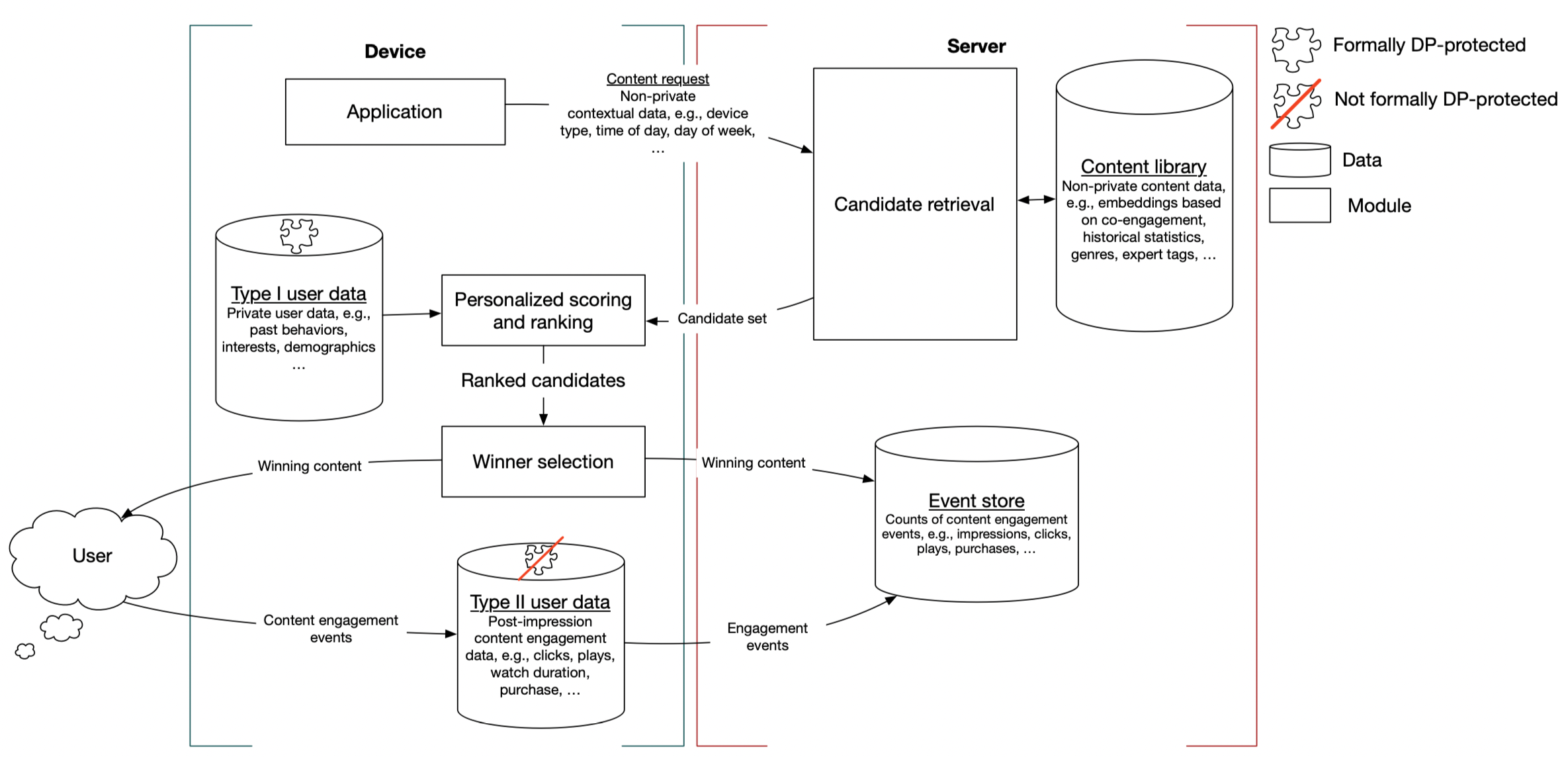}
\caption{Data flow for private, personalized recommendations.}
\label{dataflow}
\end{figure*}

\section{Background and related work}

\subsection{Recommender systems and data types}
\label{data_groups}
Recommender systems generally follows a pattern similar to that shown in Figure \ref{dataflow} \citep[see, e.g.,][]{covington2016deep, cheng2016wide, yan2021system}. First, a content request to the server initiates a lookup or candidate retrieval step, in which lower-powered, more efficient ML models perform a coarse selection of relevant candidates from a possibly very large content library. Personal data is generally not used at this stage due to the large content library size and latency requirements. After a smaller, more manageable set of candidates has been chosen from the content library, a more precise ranking step occurs, in which higher-powered models and possibly personal data are used to get more accurate predictions of engagement behaviors. Note that this step is often done at the server, but can also be on device, as is depicted. After the candidates are ranked, the top $d$ candidates (where $d$ is the number of placements available to fill) are surfaced to the user. In a non-private setting, which candidates were shown and possibly subsequent user engagement are recorded and sent to the server.

On the other hand, most approaches to private recommender systems focus on training models whose parameters do not risk leaking user data \cite{mcsherry2009differentially} but assume that the recommendation process is private: at inference time, the model will take in a user's private data and output a recommendation, which should not leave the device. However, in many real business cases, these recommendations do need to be revealed to the server. For example, an on-demand content (e.g., music, television, video, or movies) application must know how many times a piece of content was played to be able to pay the content creator appropriately. In these settings, using e.g. private federated measurement to measure noisy statistics is not legally permissible. The present work fills a previous hole in the literature---namely, that private recommender systems generally assume the recommendation phase will not be observed by the server---by outlining the key requirements we have observed in practice and presenting a system that combines existing technologies to address these needs in a novel way. Our approach, in which noise is added to the output of a private recommender system such that any recommendation is private with respect to its input data, is similar in spirit to previous private recommender systems work in which noise is added directly to the data \cite{shen2014privacy, shen2016epicrec, shin2018privacy}. 

We define two types of user data in this setting. Type I data we define as the personal data (e.g., user features such as age, gender, location, historical content engagement records, etc.) used to select what content is shown. Type II data, on the other hand, consists of any user engagement with the chosen content, including both immediate content engagement behavior, such as clicks, as well as downstream behaviors, such as play duration for a music or video platform, ad conversions for an ad platform, or purchase completion for an e-commerce platform. It is important to consider privacy protections for both types of data, as they are each potentially sensitive in nature. The algorithm we will discuss in this work provides formal privacy protections only on Type I data, so we note that additional privacy treatment would be needed to collect any Type II data. 

In addition to these two types of user data, certain non-user data is also integral to content recommendation. This data is generally not subject to privacy constraints, and can include, e.g., contextual information, such as device type, time of day or day of week; and content features, such as an item's aggregate historical engagement statistics (clicks, watches, likes, etc.), genre, expert tags, or other metadata.

\subsubsection{Key requirements}

Recommender systems are often found in two- or even three-sided marketplaces that must support the needs of both content creators and users, as well as those of the platform (e.g., revenue and compliance). Below, we identify five key requirements of a successful personalized recommendations platform:

\begin{enumerate}
    \item User privacy. We break user privacy into three distinct components, following \citep{bonawitz2021federated}:
        \begin{enumerate} 
        \item Consent/transparency: Content should only be personalized for users who have explicitly opted in.
        \item Data minimization: The amount of data persisted at central servers should be reduced to only what is needed.
        \item Data anonymization: An assurance that protected data cannot be inferred from other signals.
        \end{enumerate}
    \item User experience. Highly relevant and engaging content is recommended to users.
    \item Billing and financial compliance. Records of payment events and reasons for the payment amount are persisted. 
    \item Creator experience. Content creators can easily reach and connect with relevant audiences. 
    \item Platform revenue. The platform is able to generate enough revenue for its business.

\end{enumerate}

\subsection{Achieving user privacy in personalized recommendation}
\label{serverselection}
\subsubsection{ Consent/transparency }
Consent and transparency are achieved by obtaining explicit, opt-in consent from users before processing their data to personalize their recommendation experience. Taking advertising as an example, a prompt might appear the first time a user engages with an ad placement, asking if they would like their data to be used to personalize the ads they are shown. 

\subsubsection{ Data minimization}
Data minimization for Type I data (the features used for personalization, e.g., age, gender, location, this user's historical content engagement records, etc.) can be achieved by leaving all such data on the user's device. This data can be accessed to train ML recommendation models using federated learning techniques, such as FedAvg \citep{mcmahan2017communication}. The data will also be accessed at inference time as inputs to the production personalized recommendation model, which can be done on the device.

Similarly, data minimization can be achieved for Type II data (post-selection user engagement behaviors) by leaving the raw engagement data at the device and only collecting necessary summary statistics (e.g., the total number of clicks, downloads, etc. per piece of content).

\subsubsection{Data anonymization}

Differential privacy (DP) is becoming the de facto standard for ensuring data anonymization; among other benefits, it provides provably strong protection against several risks such as reconstruction, membership inference, and linkage attacks \citep{dwork2014algorithmic}. A mechanism $M$ is $\epsilon$-differentially private if, for any pair of adjacent databases, $D$ and $D'$, and all possible sets of outputs $S$ of $M$ \citep{dwork2011differential}:
\begin{equation}
    \label{DP}
    \frac{P(M(D) \in S)}{P(M(D') \in S)} \leq \exp(\epsilon).
\end{equation}

We note that, even if the data minimization steps described in the previous section are imposed, an honest but curious server will observe four sets of data that risk revealing Type I or Type II user data:
\begin{enumerate}
    \item Type I \& II: The trained recommendation model's parameters.
    \item Type I: The content chosen by the recommendation model. 
    \item Type I: The payment amount assessed.
    \item Type II: Any post-impression user engagement. 
\end{enumerate}

We focus on differentially-private mitigations to these risks. To begin, established methods exist to mitigate risk 1, for example, by using private federating learning variants, such as DP-SGD, in which gradients are clipped and noised before transmission to a server \citep{abadi2016deep}. Similarly, risk 4 can be mitigated by adding noise to measured aggregates, as long as they are not required for payment, e.g., following \citealt{erlingsson2014rappor}. Risk 3 is only active if payment amounts are variable and depend on Type I data; we discuss the problem of payment anonymization in more detail and introduce several ways to mitigate it specific to advertising in Section \ref{pricing}. However, risk 2 is more problematic---in settings where payments are involved, exact records of payment events are legally required so that content creators (advertisers) can be paid (charged) the correct and fair amount. In the most privacy-friendly settings, we can assume that payment events are the impression of a piece of content;\footnote{Though payment events in practice are often downstream engagement metrics such as clicks or conversions, i.e. Type II data.} in this case, DP noise should not be added to the total impression counts of each piece of content, as might otherwise be the obvious approach, leaving Type I data at risk of being revealed.

In this work, we propose addressing the data anonymization risk described above (risk 2) by pulling DP noise addition forward from measurement into the content selection process. In particular, a differentially-private, randomized algorithm can be used to select a piece of content given a set of personalized (private) scores. Rather than greedily selecting the content with the highest personalized score, the choice can be drawn from a distribution satisfying differential privacy. Interestingly, such an approach has strong overlap with the notion of ``exploration'' in the bandits literature; approaches that introduce stochasticity into the recommendation process for the sake of privacy will simultaneously achieve certain exploration goals. In the next section, we discuss the similarities between privacy and exploration in more detail and describe how established differentially-private mechanisms can be applied in the personalized recommendation setting. 


\subsection{Randomized algorithms for differentially-private recommendations}
Below we describe two existing approaches to differentially private measurement that we will explore for use in our setting. Note that these approaches could practically be substituted with alternative private measurement algorithms. 
\label{dpselection}
\subsubsection{Randomized Response}
Randomized Response (RR) is one of the oldest differentially-private techniques in the literature, dating back to surveys conducted by Warner \citeyearpar{warner1965randomized}. The algorithm instructs survey respondents to flip a weighted coin before answering a potentially sensitive question. If the coin returns heads, they should answer the question truthfully. However, if the coin returns tails, they should select a response uniformly at random from those available. Originally developed for the binary-choice survey setting, it can be easily extended to the personalized recommendation setting where there are $a$ choices (e.g., $a$ pieces of content to choose from) and the ``truth'' is the choice with the highest private score.  A piece of content with private score, $s_i$, is drawn from the following probability distribution:

\begin{equation}
\label{equation-rr}
p_i = 
\begin{dcases}
    \frac{e^{\epsilon}}{a-1+e^{\epsilon}} & \text{if } s_i=argmax_i s_i\\
    \frac{1}{a-1+e^{\epsilon}}              & \text{otherwise.}
\end{dcases}
\end{equation}

We observe that this algorithm is equivalent to ``epsilon-greedy'' in the bandits literature, named as such because with probability $(1-\epsilon_{RL})$ the algorithm ``greedily'' chooses the highest scoring item from a set, while with probability $\epsilon_{RL}$ it explores, choosing an ad from the set uniformly at random.  The link between RR and epsilon-greedy has been noted previously \citep{hannun2019privacy}. Note the clash of notations; from here on, we will use $\epsilon$ exclusively to refer to the privacy parameter of differential privacy.

\subsubsection{Report (Select) Noisy Max}

Another common set of $\epsilon$ differentially-private randomized algorithms are the ``Report Noisy Max'' family (Algorithm \ref{rnmg}). These algorithms generally follow the format in which noise drawn independently from a random distribution is added to each choice's private score to achieve a noisy score. Then, the choice with the highest noisy score is reported. Often, Report Noisy Max is applied in central DP problems: for example, if one wishes to know the most common disease in a healthcare database, Report Noisy Max would entail 1) counting the number of instances of each disease, 2) adding noise to these counts and 3) reporting the disease whose noisy count was highest to the analyst. In our case, we will use these algorithms to provide local (rather than central) $\epsilon$-DP, and we will not ``report'' but rather ``select'' the noisy max. 
\begin{algorithm}
\caption{Select Noisy Max}
\label{rnmg}
    \begin{algorithmic}
    \label{snm}
    \STATE \textbf{Input:} Set of choices, $A$, with private scores, $s$, and sensitivity, $\Delta$
    \FOR{content $i$ in set $A$}
    \STATE $s_{i,noisy}=s_i+N(\frac{\epsilon}{2\Delta})$
    \ENDFOR
    \STATE \textbf{Return:} $argmax_i s_{i, noisy}$
    \end{algorithmic}
\end{algorithm}

Relative to RR, Select Noisy Max (SNM) algorithms possess an additional parameter: the sensitivity, $\Delta$, defined as the maximum possible amount that private information can alter any score. When the private scores have high variability, the sensitivity can be reduced in several ways. One possibility is to scale the private scores of the candidates per decision, e.g.:

\begin{equation}
s_{scaled} = \frac{(s - min(s) )}{ (max(s) - min(s))}.
\end{equation}

However, this approach may be suboptimal if informative non-private scores are available for each candidate. In this case, the amount by which the private scores can change relative to the public scores may be clipped. When this is done, the private score used by the algorithm is the clipped score:

\begin{equation}
s_{clipped} = \max\left(\min\left(s_{server}+\frac{\Delta}{2}, s\right), s_{server}-\frac{\Delta}{2}\right),
\end{equation}

where $s_{server}$ is the score without private information. While efficient in the presence of reasonably predictive non-private scores and low $\epsilon$, this approach has the undesirable quality of not converging to the non-private limit as $\epsilon \rightarrow \infty$.  

Common choices for $N$, the noise distribution, include Gumbel, Exponential, and Laplace noise. When Gumbel noise is added, the probability distribution has a simple formula:
\begin{equation}
    p_i=\frac{exp(s_i \times \frac{\epsilon}{2\Delta})}{\sum_j exp(s_j \times \frac{\epsilon}{2\Delta})}.
\end{equation}
This variant is also known as the Exponential Mechanism \citep{mcsherry2007mechanism}, and is equivalent to ``soft-max'' or ``Boltzmann'' exploration in the bandits and reinforcement learning literature (see, e.g., Sutton and Barto \citeyear{sutton2018reinforcement}). Recently, it was shown that Report Noisy Max with exponential noise has a higher expected utility than the exponential mechanism for the same privacy parameter $\epsilon$ \citep{ding2021permute, mckenna2020permute}. Therefore, we focus on exponentially-noised SNM in our experiments.

\section{Algorithm}
\label{algorithm}

\begin{algorithm}[tb]
   \caption{Differentially-private, personalized recommendation.}
   \label{alg:boas}

\begin{algorithmic}
   \STATE \textbf{INPUT:}  Server (non-private) context $c_{server}$, device (private) context $c_{device}$. Parameters: score cutoff, $\gamma$; privacy parameter, $\epsilon$; (possibly) clipping bound, $\frac{\Delta}{2}$.
   \STATE {\bfseries Server executes:} 
   \FOR{each eligible candidate in $r_1...r_k...r_K$}
   \STATE $s^{server}_k$ $\leftarrow$ getServerScore($r_k$, $c_{server}$)
   \ENDFOR
   \STATE $\overrightarrow{P}$ $\leftarrow$ getPrices($\overrightarrow{S^{server}}$)
   \STATE $r_1...r_b...r_B \leftarrow$ getFinalCandidates($\overrightarrow{S^{server}}$, $\gamma$)
   \STATE Return $r_1...r_b...r_B$ to device
   \STATE {\bfseries Client executes:}
   \FOR{each candidate in $r_1...r_b...r_B$}
   \STATE $s^{device}_b$ $\leftarrow$ getPrivateScore($r_b$, $c_{server}$, $c_{device}$)
   \ENDFOR
   \STATE $r_{chosen} \leftarrow$ randomSelect($\overrightarrow{S^{device}}$, $\epsilon$, $\frac{\Delta}{2}$) // e.g., RR or SNM
   \STATE {\bfseries Server observes:} $r_{chosen}$ and any engagement events, $E$
   \STATE {\bfseries Server pays or charges:}  $p_{chosen}$
   \STATE 
   \STATE {\bfseries getFinalCandidates}($\overrightarrow{S^{server}}$, $\gamma$):
   \STATE candidates = [] // Initialize empty set
   \FOR{$s_i$ in $\overrightarrow{S^{server}}$}
   \IF{$s_i \geq max(\overrightarrow{S^{server}}) \times (1-\gamma)$}
   \STATE candidates+=$r_i$
   \ENDIF
   \ENDFOR
   \RETURN candidates
   \STATE 
   \STATE 

\end{algorithmic}
\end{algorithm}

Our proposed algorithm is applicable to any recommender system setting in which the recommended content must be recorded precisely at the server without the risk of exposing Type I user data. We desire an approach that guarantees user privacy (requirement 1) and financial compliance (requirement 4), without sacrificing too much utility (requirements 2, 4, and 5); our algorithm guarantees (1) and (4) and we test its utility in our experiments. At a high level, we combine a recommender system model trained privately (e.g. using DP-FedAvg) with a private selection algorithm (e.g. RR or SNM) to yield recommendations that can be observed precisely by the server while protecting Type I user data. Note that we do not provide an explicit theoretical analysis of our method since it is simply a combination of previously published private algorithms.The full approach is detailed in Algorithm \ref{alg:boas}. Below we describe each step in detail: 
 \begin{enumerate}

 \item Server-side scoring and candidates retrieval. Initially, a recommendation opportunity (e.g., the user navigating to a page on their device) begins the process. At the server, all eligible candidates are scored using the non-private context, $c_{server}$. The non-private context is any non-user-specific information that the server has to help in the first stage content retrieval process. For example, a search term, the local environment (such as what else is on the page, where on the page this content would be shown, what song is currently playing, etc), contextual knowledge about time of day, day of week, time of year, or knowledge about what is currently trending. Additionally, there is non-private information about each piece of content specifically, such as its genre, expert tags or labels, and historical population-wide engagement statistics (e.g., click rate, watch rate, skip rate, etc.).

 \item Server-side ranking. Candidates are ranked according to their score, $s^{server}$. If payment amounts can depend on the context (as in advertising), a set of non-private prices, $\overrightarrow{P}$, which depend on $\overrightarrow{S^{server}}$, may also be calculated and persisted (more details in Appendix \ref{pricing}).

 \item Server-side final candidate selection. Next, the full set of eligible candidates is optionally reduced to a smaller set of final candidates to be sent to the device. This step is important for managing communication costs; too large of set sizes will result in dropped requests and time-outs. As we will see in Section \ref{results}, it can also have a benefit for privacy in approaches where the amount of noise required is proportional to the signal dimensionality, like randomized response. Our approach for optionally reducing the set size is fairly straightforward. We first identify the highest scoring candidate, $max(\overrightarrow{S^{server}})$ , where $\overrightarrow{S^{server}}:= [s_{server, 1}, \dots, s_{server, k}, \dots,s_{server, K} ]$. Then, to ensure candidates with sufficiently high scores are sent to the device, we define a cutoff parameter, $\gamma$, and send only candidates with $s_{server}$ higher than $(1-\gamma) \times max(\overrightarrow{S^{server}})$. This cutoff parameter, $\gamma$, significantly impacts the different metrics considered in this paper and is studied explicitly in Section \ref{results}.

 \item On-device winner selection. When the final set of candidates reaches the device, the device calls getPrivateScore to get private scores for each candidate. Finally, a random algorithm (randomSelect), which could be, e.g., RR or SNM chooses the content to show based on the private scores. 

\item Payment. At the end of this process, the server observes (without noise) the selected content and pays (charges) the content creator (advertiser) appropriately. 
 
\end{enumerate}

The benefits of the above approach are that the output of a model that consumes private, on-device user data (Type I), can be observed directly with DP guarantees protecting its private inputs. The primary drawback is the addition of noise during recommendation will necessarily lower utility (i.e., user, creator and platform experiences). In the next section, we empirically evaluate the utility cost of the proposed algorithm compared to both a non-private and non-personalized recommendation setting.


\section{Evaluation}
\label{Evaluation}

 \subsection{Methods}

To evaluate our proposed algorithm, we consider the digital advertising setting. In particular, we assume a second-price auction format in which advertisers specify their bids per click, $b_{click}$, and ads are scored according to their effective bid per impression opportunity, $b_{imp}$. Thus, the scores in this setting are $s^{server}=b^{server}_{imp}= bid_{click} \times pClick_{server}$ and $s^{device}=b^{device}_{imp}= bid_{click} \times pClick_{device}$, where $pClick_{device}$ and $pClick_{server}$ are the modeled probabilities of an impression on this ad resulting in a click, conditioned on either both device (private) and server data or only server data, respectively. Note that, while this choice for the score, $s$, is standard in digital auctions, $s$ could easily be a different metric, e.g., one that accounts for long-term revenue effects or simply a quality score in a pure recommendation setting.

In the second-price auction format, the payment amounts are context-dependent. We choose to remove private information from pricing to avoid privacy leakage via price observation (Appendix \ref{pricing}). Therefore, the price each candidate pays if it wins the impression is calculated as  $price_r=bid_{r+1} \times pClick_{r+1, server}$, where the subscript $r$ represents the candidate's server (non-private) rank.

We conduct our experiments on two datasets, a public Alibaba dataset and an internal dataset derived from our search ads auctions. Details for these datasets are given in Appendix \ref{dataset_details}.

To evaluate our approach, we analyze click-through rate to measure user experience (requirement 2), surplus to measure advertiser experience (requirement 4), and revenue to measure platform experience (requirement 5). In general, we must approximate surplus because we do not know advertisers' truthful valuations for clicks. In the public Alibaba dataset, in which the advertiser's bids are not included as part of the data release, we assume that advertiser's click valuations are proportional to the price of the item being advertised. In the internal dataset, where we do have bids, we assume that advertisers are bidding ``truthfully,'' which is a dominant strategy in the second-price auction format. Additionally, in the internal dataset, we do not know counterfactual click values and thus must approximate them; to do so, we follow the ``direct method'' and use $pClick_{device}$ to estimate clicks. Therefore, our metric definitions are as follows:
\begin{itemize}
    \item Click-through-rate (CTR): $\frac{m}{n} \approx \sum_{i=1}^{n} pClick_{device, i}$,
    \item Advertiser surplus: $\sum_{i=1}^{m}(v_i-p_i) \approx \sum_{i=1}^{n}(b_i \times pClick_{device, i} - p_i )$,
    \item Platform revenue: $\sum_{i=1}^{n}p_i$,
\end{itemize}
where $m$ in the total number of clicks, $n$ is the total number of impressions (which we consider to be equal to the number of ad requests/opportunities for the purpose of this work), $p_i$ is the price charged for an impression to the $i^{th}$ ad, $v_i$ is the amount that the $i^{th}$ advertiser values an impression on their ad, $pClick_{i, device}$ is the estimated probability of clicking for the internal data, and the real (0/1) logged click for the Taobao dataset, and $b_i$ is the real click bid for the internal dataset and the item price times a proportionality constant for the public dataset. We note that these are offline simulations only, and we ignore any monetary advertiser budget constraints; i.e., advertisers are assumed to have infinite budget. Code to reproduce all experiments is available at: REDACTED. 

\begin{figure}[t]
\centering
\includegraphics[width=\columnwidth]{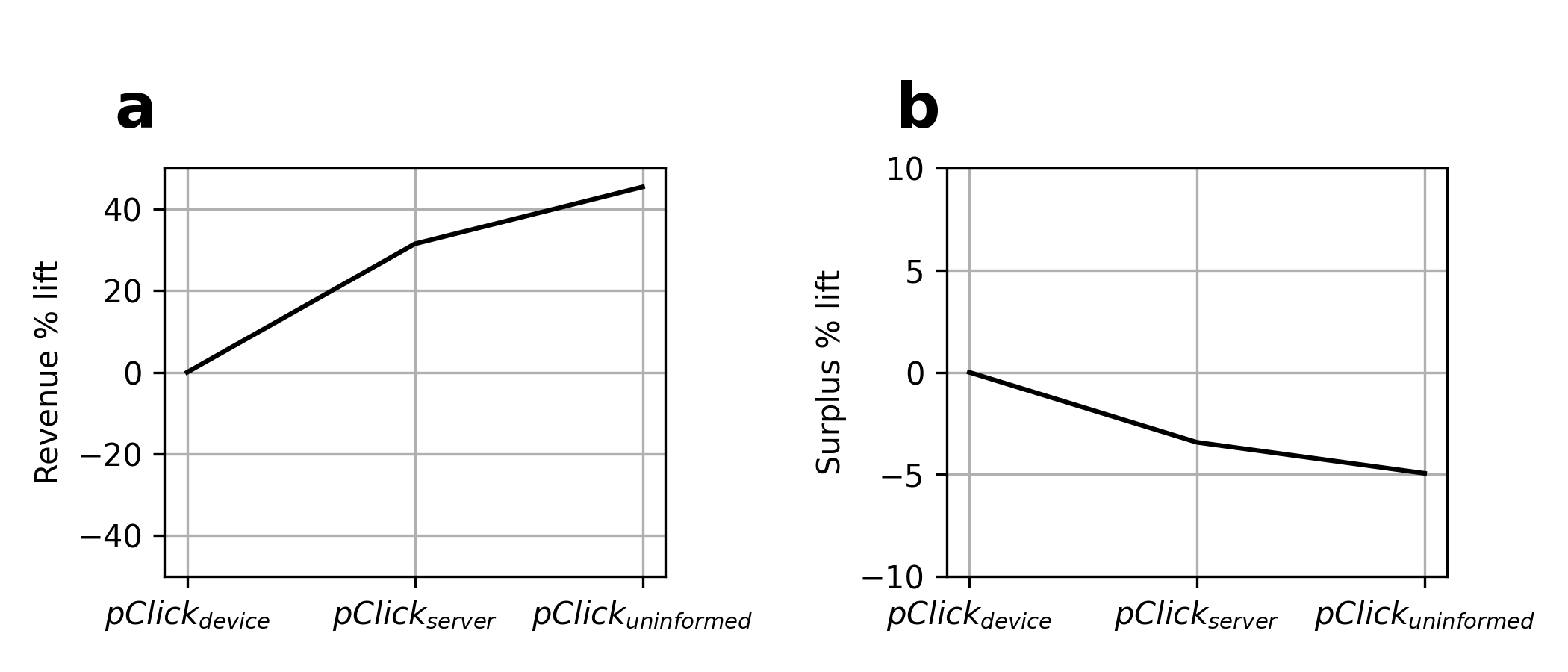}
\caption{Effects of removing information from the $pClick$ model on the internal datset. Lifts are relative to the full-information ($\epsilon=\infty$) auction setting. }
\label{fig-pricing}
\end{figure}

\begin{figure*}[t]
\centering
\includegraphics[width=\textwidth]{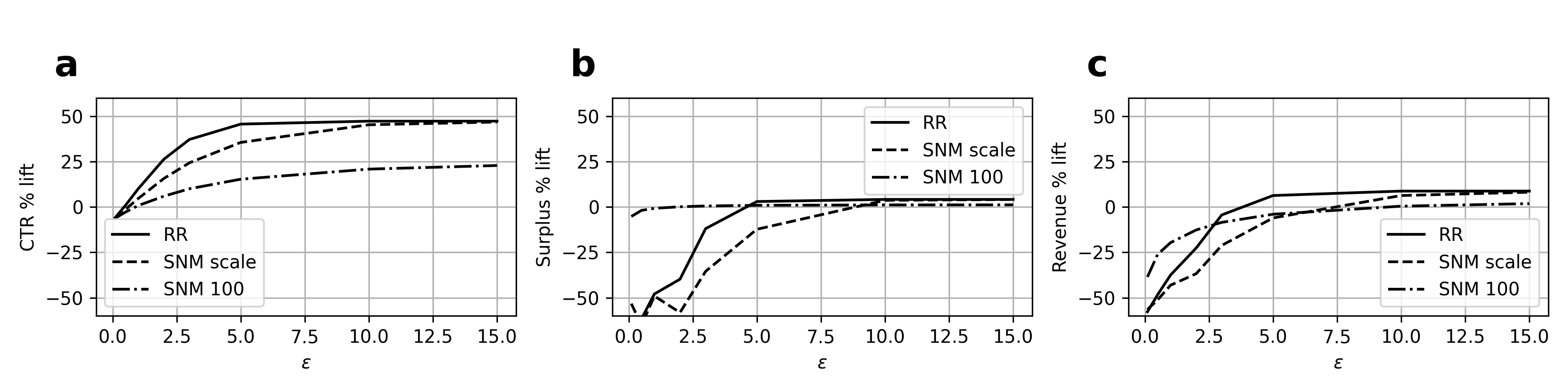}
\caption{Effects of the privacy parameter, $\epsilon$ on (a) CTR, (b) surplus and (c) revenue for SNM with either scaling or clipping bound = 100 and RR on the internal datset. (d)-(f) show the effects of changing the clipping bound in SNM.}
\label{rnm-vs-rr}
\end{figure*}

\subsection{Research questions}
 Our goal in the proceeding experiments is to gain insight into the following questions:
\begin{enumerate}
    \item What are the effects of removing private information from pricing?
    \item How sensitive are our key metrics to the privacy parameter, $\epsilon$, and how does the choice of randomized algorithm (RR and variants of SNM) affect the results?
    \item What is the effect of decreasing the number of choices sent to device via the cutoff, $\gamma$?
\end{enumerate}

 \subsection{Results}
\label{results}

\subsubsection{Effects of removing private information from pricing}
\label{results-pricing}
First, we examine the effects of removing personalized user information from pricing. To isolate the pricing effects from differences in ad selection, we use a greedy (non-private) selection mechanism and set $\gamma=1$, such that the best candidate will always be chosen (CTR is unaffected). We then compare the setting in which $pClick_{device}$ is used initially for pricing to one where $pClick_{server}$ is used. We observe (Figure \ref{fig-pricing}) a large increase in platform revenue relative to the case where private data is used to price, which comes at a direct cost to advertisers (lower surplus). That removing information from the $pClick$ model would have such a large effect on revenue and surplus initially surprised us, so we conducted an additional experiment, in which an uninformed $pClick$ model (that predicts the global CTR average for each candidate) is used. Indeed, we observe the trend continue for the naive model. In fact, this observation is consistent with previous works \citep{levin2010online, hummel2016does, de2016online, bergemann2011targeting} and discussed further in Appendix \ref{pclick-accuracy-pricing}.

\subsubsection{Comparison of RR and SNM}
\label{results-snm}
Next, we examine the effects of using RR or SNM as a function of the local privacy parameter, $\epsilon$, and, if applicable, the clipping bound (we consider both the ``clipped'' and the ``scaled'' variants of SNM). Our results are shown in Figure \ref{rnm-vs-rr}. We see that, for low values of $\epsilon$, SNM with clipping can outperform RR. However, at these low values of $\epsilon$, neither mechanism succeeds in outperforming the un-personalized baseline (the lift is negative). At higher values of $\epsilon$ (less privacy), we see that the trend shifts and RR begins to outperform SNM (both clipping and scaling). Notably, SNM with clipping plateaus before reaching the no-privacy limit because the clipping mechanism bounds the effects of private information. The dominance of RR over SNM surprised us, so we analyzed their performance on smaller samples of real and synthetic scores in Appendix \ref{snm-vs-rr}. We find that in cases where a) the private score distribution is extreme, b) more candidates are available, and c) informative public scores are available, SNM can outperform RR. Thus, we note that the mechanisms' relative performance is likely to be quite data-dependent.

Given these results, we conclude that purely local DP mechanisms may not have sufficient utility for private, personalized recommendation. The amounts of noise at e.g., $\epsilon=1$ are so high that our metrics are worse than the non-personalized baseline. However, pairing this technique with a privacy amplification approach could facilitate the use of higher local $\epsilon$ values; in particular, if impressions are shuffled before they are returned to the server, the server would retain the needed information of precisely how many times an ad was impressed without being able to attribute individual ad impressions to any particular user (more details in Appendix \ref{amplification}). In this regime, RR appears to be the more appropriate randomized algorithm. Choosing RR with $\epsilon=5$, for example, yields significant increases in metrics relative to the unpersonalized baseline and approaches the performance of the fully personalized (non-private) case. Therefore, we conclude that applying RR with intermediate $\epsilon$ values in a setting that ensures only aggregated (shuffled) impression reports are released could provide a sufficient privacy-utility trade-off for the business. Additionally, we note that this analysis ignores the benefits of non-deterministic content selection in subsequent model training, where it may, e.g., be an effective counter against selection bias and covariate shift.

\subsubsection{Effects of decreasing content set size}
\label{results-cutoff}
\begin{figure*}[t]
\centering
\includegraphics[width=\textwidth]{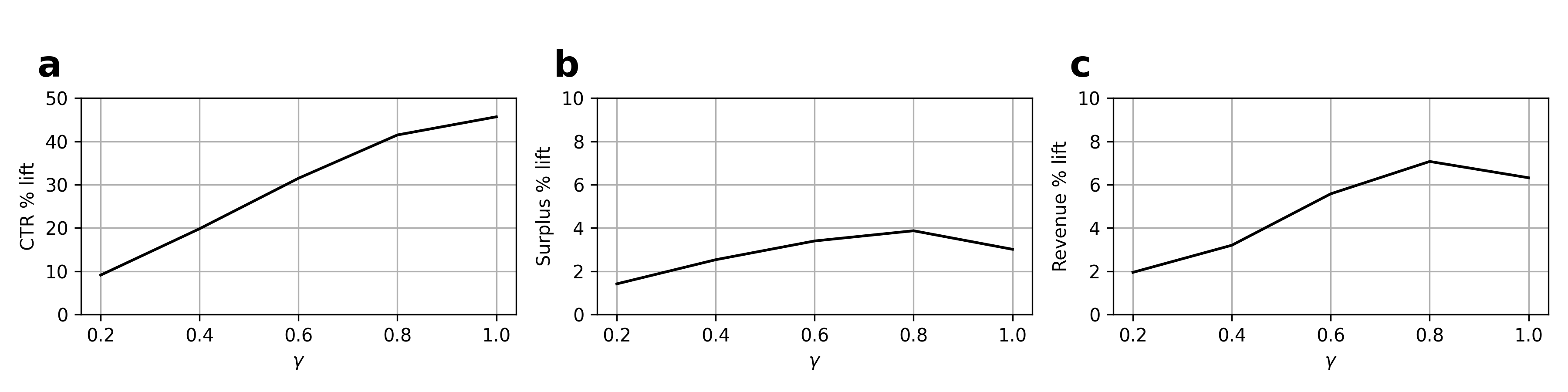}
\caption{Effects of the changing the cutoff parameter, $\gamma$, on (a) CTR, (b) surplus and (c) revenue  on the internal datset.}
\label{rr-cutoff}
\end{figure*}

We next examine the effects of varying the number of candidates sent to device using RR with $\epsilon=5$, which we enforce by the cutoff parameter, $\gamma$, as in Algorithm \ref{alg:boas}. We observe (Figure \ref{rr-cutoff}) that the metrics pass through an optimum at an intermediate value near $\gamma=0.8$, which we justify as a trade-off between recall and privacy. At low cutoff parameters (smaller sets), the process is limited by recall---the set of final candidates sent to the device is less likely to contain the best candidate given the private data. At high cutoff parameters (larger sets), the process is limited by the decreasing probability density on the greedy candidate in the RR mechanism (Equation \ref{equation-rr}).

\subsubsection{Results on internal data}
We also conducted our experiments on an internal dataset from our search ads business (see Appendix Figures \ref{internal-pricing}, \ref{internal-rnm-vs-rr}, \ref{internal-rr-cutoff}). We see similar results with a few exceptions. First, the trend in pricing remains: the less accurate the $pClick$ model, the higher the prices (Figure \ref{internal-pricing}). Second, RR still outperforms SNM at higher values of $\epsilon$ (Figure \ref{internal-rnm-vs-rr}(a-c)). The main difference is in the CTR metric, for which SNM was never dominant in the Alibaba dataset. In the internal dataset, we now see a crossover in RR vs SNM performance for CTR as well as revenue and surplus. We explore the differences in more detail in Appendix \ref{taobao-vs-internal-results}.

\section{Conclusion}

To summarize, we find that precise measurement of impression events for payment in personalized recommendation settings is possible while maintaining strong differential privacy guarantees on user data. The randomized algorithms we considered can achieve reasonable utility at intermediate levels of epsilon (e.g., $\epsilon=5$). Paired with a shuffling approach to ensure that impression events cannot be linked to individual users, the total privacy loss of the system could be reduced while meeting the business need of precise measurement in aggregate. For the datasets we studied, RR was found to outperform SNM at relevant privacy levels. However, we also demonstrated (Appendix \ref{snm-vs-rr}) that their relative performance is situational, depending on the private score distribution, the presence of informative public scores, and the number of choices available in the personalized ranking stage. Therefore, we recommend considering both algorithms when approaching a new application.

\section{Limitations and future work}

One limitation of this work is that many platforms today support Type II data payment events, rather than the simple case considered here of a pay-per-impression model. For example, advertising businesses may only charge advertisers when a click or conversion is actually observed, rather than charging per ad impression. Our method of adding noise to the selection process only provides differential privacy to Type I data (the data used as input to personalized recommendation models), not to Type II data (how users react to the content they are shown); if payment events are also Type II data, it is not clear how they could be measured both precisely and privately. How this discrepancy might be addressed is an important area for future study.

Additionally, the two exploration algorithms considered in this work select recommendations based only on the point estimates of their private scores. The bandits literature goes far beyond this relatively naive approach to regret minimization. Instead of selecting based on the point estimates alone, it is often more effective to include the uncertainty in the estimates as well. For example, a choice with just a few previous observations compared to a choice with thousands of previous observations might be worth exploring more, even if they both have the same predicted utility. Therefore, DP variants of techniques like upper confidence bound \citep{lai1985asymptotically, auer2002finite, agrawal1995sample} and Thompson sampling \citep{thompson1933likelihood} would be worthwhile to explore in the future. 

\section{Acknowledgements}
The authors gratefully acknowledge the help of Audra McMillan, Daniela Antonova, Barry Theobald, Jonathan Whitmore, Jessica Zuniga, Robert Monarch, Kanishka Bhaduri, and David Betancourt in providing comments on early drafts of the paper and helping to shape its final direction. We further acknowledge Audra McMillan for her guidance and support throughout the submission process. Early idea-shaping discussions with Kunal Talwar, Brian Knott, Luke Winstrom, and Michael Scaria were also helpful.


\bibliography{main}


\appendix
\onecolumn

\section{Differentially-private approaches to pricing in advertising}
\label{pricing}

\label{pricing-intro}

In certain settings, the payment amount is dependent on the private scores and context. For example, in the many digital advertising settings where ``truthful'' auction mechanisms are employed (e.g., second-price and Vickrey-Clarke-Groves), the price charged for an ad will depend on the other candidates and their scores. Specifically, in a second price auction in which advertisers specify their bids per click, the price an ad will pay if a click is observed (in a pay per click model) is $price_r=bid_{r+1} \times \frac{pClick_{r+1}}{pClick_{r}}$, or $price_r=bid_{r+1} \times pClick_{r+1}$ if an impression is observed in a pay-per-impression model, where the subscript represents candidates' rank. $pClick_r$ is the probability of a click on the $r^{th}$ candidate. The $pClick$ terms arise because advertisers are bidding on clicks, which are events downstream of the platform's decision: which ad to show in an open ad slot. If the ad has no lower ranked ads below it, it will pay a ``reserve'' price, specified by the auction platform. Ads with bids lower than the reserve price will not be eligible for selection. 

Clearly, $pClick_r$ is influenced by private user data and poses a risk of data leakage, and thus the price also poses this risk. Additionally, the price will contain information regarding the ordering of candidates (i.e., which candidate was the price setter) through the $bid_{r+1}$ term, which is known by the ad server and transmitted to the device for use is private ranking. Therefore, if the payment amount in such a setting is revealed in the clear, there is significant risk of privacy loss. While privacy risks related to payment will be application-specific, we considered several possibilities for approaches to private pricing in the advertising setting:
\begin{enumerate}
    \item First-price auction mechanisms
    \item Price using noisy $pClick$ values
    \item Price without private data
\end{enumerate} 
We choose to use approach 3 in our algorithm (price without private data), and we detail the reasons for our choice in the following subsections.

\subsection{First-price auction mechanisms}
Since 2018, major display ad auctions are shifting from second-price auction to first-price auctions. There are several reasons for this wider adoption of first-price auction, including the widespread growth of header bidding, and increased demand for transparency and accountability \citep{hovaness2018}. The first-price auction format has a unique advantage when it comes to protecting user privacy: the price charged for an observed click is always equal to the winner's bid, mitigating the risk of information leakage. However, the downside to first-price auctions is well-known in the literature; first-price auctions encourage ``non-truthful'' bidding behaviors, in which advertisers' bids do not reflect their values \citep{roughgarden2016twenty}. Additionally, first-price auctions incentivize bidder to shade their bids to improve their returns \citep{zhou2021efficient,pan2020bid,zhang2021meow}, which could lead to unstable marketplaces \citep{edelman2007strategic}. Thus, to avoid bidders' strategical behaviors such as bid shading and maintain market stability, we choose not to use first-price auction in our algorithm.

\subsection{Price using noisy pClick values}
Another option to anonymize ad prices is to use noisy $pClick$ values during pricing. The SNM family of ad selection algorithms lend themselves to this approach, as the selection mechanism relies on noisy scores. However, by Jensen's inequality, the ratio of two independently noised quantitites will be biased, leading to higher prices paid on average. Additionally, revealing the noisy score would incur additional privacy loss, and for algorithms that do not rely on noising the private score, such as RR, private pricing would require a separate randomized algorithm.

\subsection{Price without private data}
In the final option we consider, ads could be ranked and priced based on non-private data only. This ranking would not be used to select the final ad (ad selection would be based on private scores), but the winning ad (if clicked) would be charged the price from the non-private score ranking. We opt to use this approach (pricing without private data) in our algorithm, as it does not have the downsides of unsteady marketplace equilibria or price bias. In Section \ref{results}, we explore the implications of this choice on pricing.

\section{Privacy amplification}
\label{amplification}
Note that if a strong enough local differential privacy level is chosen, then the selected content can be measured directly and linked to its originating content request and user, minimally affecting the overall platform ecosystem. However, requiring strong local DP guarantees can significantly degrade user, creator, and platform experience by reducing marketplace efficiency. The requirement of precise measurement for fair payment means only that aggregate statistics, i.e., precisely how many times a piece of content was shown, not individual impression events, are necessary. Therefore, it is possible to increase the utility of the process while maintaining privacy by applying a model for differential privacy that ensures only aggregated or anonymized reports are released. In such models, the small amount of noise applied in the local process (normally during measurement, but in our case during content selection itself), can be subsequently amplified during the aggregate measurement process via shuffling \citep{DBLP:journals/corr/abs-1808-01394, DBLP:journals/corr/abs-1811-12469}. For example, a Prio-like secure aggregation system could provide aggregate statistics \citep{corrigan2017prio} with higher overall DP guarantees by masking which device sent which impression event without adding more noise. Thus, in our setting, the measured impression counts of each piece of content would still be precise. 

\section{Regret bounds}
\label{regret}
It has been previously proven that differentially-private contextual multi-armed bandits will have linear regret if they must protect the privacy of the present action being taken (due to the dependence of the best action on the private context) \citep{shariff2018differentially}. A notion of ``joint'' differential privacy was proposed to resolve this that achieves sub-linear regret bounds, but we note that this relaxation does not fit our use case, in which a third party (e.g. the platform) will observe the output of the algorithm; it only works to protect one user's personal data from leakage to other users through the recommendations they see. The intuition is straightforward: a DP requirement on the selected action will mean that there is a minimum probability density enforced on each action for all time---contrary to the typical assumption in bandits literature, gaining more certainty about a ``bad'' action's quality will not always mean being able to reduce its selection probability in the private setting.  

\section{Digital ads overview}
\label{ads}
Search ads placements are generally keyword-based auctions, in which advertisers can bid for specific search terms. When a user issues a search from their device, the ad platform retrieves the set of all advertisers actively bidding on this (and possibly other related) search terms. Next, relevance models are used to evaluate each ad for its relevance, and any advertisers bidding on search terms not relevant to their product are removed. Finally, an auction is run to select the final ad (or possibly ads if multiple slots exist) to show. In general, the auction ranks eligible ads by the amount they are willing to pay for their ad to be shown in the available slot (possibly also mixing in predictors of long-term revenue). In a bid-per-impression auction, this means ads are simply ranked by their bids. However, many auctions are bid-per-click or even bid-per-conversion. In these cases, bids must be converted to units of dollars per impression using predictive models, either $pClick$ (probability of click given an impression) or $pConv$ (probability of conversion given an impression) models. Finally, the top $d$-ranked ads are selected to be shown in the $d$ available ad slots. We focus on bid-per-click single-slot ads auctions, meaning the ad with the highest product of bid and $pClick$ is the winner. If the ad is shown, the advertiser will be charged. The amount they are charged is dependent on the pricing mechanism (e.g., first or second price). Pricing is discussed in more details in Section \ref{pricing}.

\section{Dataset details}
\label{dataset_details}

\subsection{Alibaba (Taobabo) dataset}
\label{taobao_details}

The Alibaba Dataset is a public dataset collected from the online display advertising traffic logs in Alibaba \citep{tianchi2018}. A total of 1,140,000 users are sampled from the website of Taobao for 8 days of ad display / click logs with a total of 26 million records. 

We use the following 3 tables: raw\_sample, ad\_features, and user\_profile. The raw\_sample table contains user-ads interaction records, including user ID, timestamp, adgroup ID, scenario ID and click log. The ad\_features table includes information for each ad, such as the item it is advertising, its category and its price. The user\_profile table contains user-related features like gender information, age level, consumption grade, shopping level, student tag, city level wealth tag, etc. This dataset is typically used to predict the probability of clicking on an ad when impressed based on user’s history shopping behavior (CTR / pClick prediction), as in \citep{zhou2018deep, gai2017learning}.

The Taobao dataset does not naturally have auction candidates, bids, or pClick values. To get $pClick_{device}$, we use the model from \citep{zhou2018deep} and consider all user features to be private (i.e., only available on the device). To get $pClick_{server}$, we train a FNN model  that does not include user features. We use the DeepCTR library for training the pClick models \citep{shen2017deepctr}.  To simulate ads auctions, we assume that all ads with identical userid and timestamp are candidates in a single-slot auction. To approximate bids, we assume that the prices of the items being advertised are proportional to the amount the advertisers would value a click on their ad, and thus set click-bids to be proportional to the prices of the advertised items. Finally, we set the reserve price equal to the minimum bid.

\subsection{Internal data}

The internal dataset is derived from real search ads auctions logs. Our auctions are a second price, bid-per-click or bid-per-conversion auction format.\footnote{For more details on search advertising, see Appendix \ref{ads}.} We sampled approximately 750k auction logs from a two week period in November 2022.  For each auction, we collected the set of advertisers who were active in that auction, and we artificially limited the set to the 15 top-ranked candidates. For each advertiser, we collected their click-bid (the maximum amount they specified they were willing to pay per click) and two modeled pClick estimates. The first pClick estimate, which we use for $pClick_{server}$ in our simulations, has only ad and context features (e.g., the query, the country, and the historical statistics for this ad type, advertiser, creative, etc); the second pClick estimate, which we use for $pClick_{device}$, also includes several coarse demographic features of the user, such as their age range and gender. These coarse user features are available at the server in our current production system for users who have opted into personalized advertising, but we assume in our simulations that we would want to store and process them privately on-device.  

\section{Effects of pClick model accuracy on pricing in second-price pay-per-click ads auctions}
\label{pclick-accuracy-pricing}
A review of the existing literature in this area generally supports the finding that increased targeting in advertising can lower equilibrium prices \citep{levin2010online, hummel2016does, de2016online, bergemann2011targeting}.\footnote{Conditional on a static reserve price; \citep{sayedi2018real} demonstrate that these effects can often be negated by appropriate reserve price modifications.} This effect is attributed to market thinning---for example, in the extreme case in which each consumer will buy only one item and each advertiser has complete information, there would be only one bidding advertiser per consumer, and thus no competition or price support \citep{hummel2016does}. To extend from targeting to the present case of pClick accuracy, Rafieian and Yoganarasimhan \citeyear{rafieian2021targeting} prove that they are in fact equivalent, with the control over targeting granularity simply moving from the advertiser to the platform. As previously noted \citep{de2016online, rafieian2021targeting}, this result suggests that advertising platform incentives may be aligned with user privacy. However, we note that it is also directly at odds with advertiser incentives who will experience lower return on ad spend.

\section{Deep dive into relative performance of RR and SNM}
\label{snm-vs-rr}

\begin{figure}[ht]
\vskip 0.2in
\begin{center}
\includegraphics[width=\columnwidth]{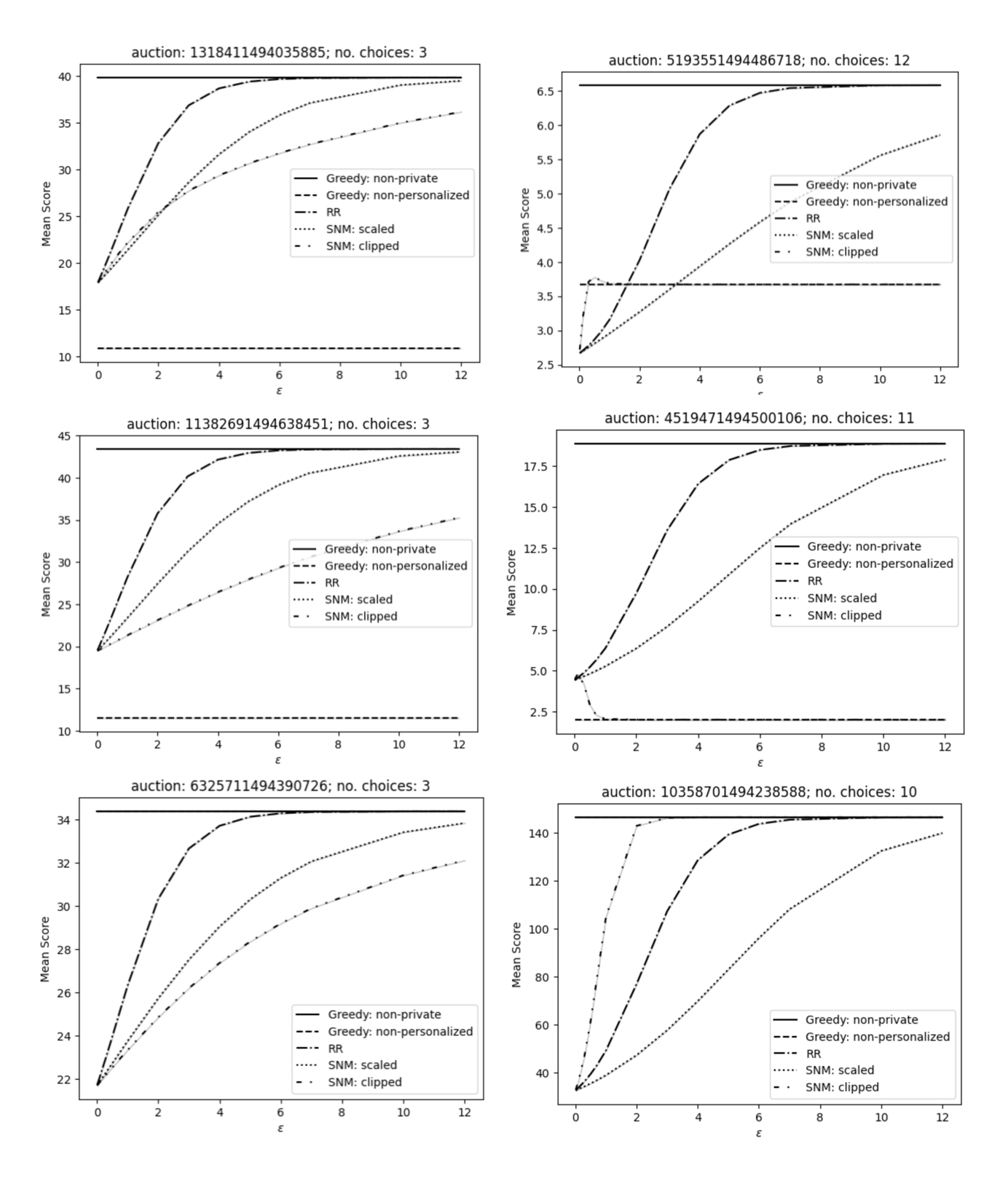}
\caption{Comparison of the performance of RR, scaled SNM with Gumbel noise, and clipped SNM with Gumbel noise on a set of samples from the Taobao dataset. The auction ids are the concatenation of the timestamp and the user ID in the original dataset. The no. choices are the number of candidates available to choose from. The mean score is the value of each mechanism in expectation.}
\label{taobao-samples}
\end{center}
\vskip -0.2in
\end{figure}

In Figure \ref{taobao-samples}, we sample 6 auctions from the Alibaba dataset and show the relative performance (expected values using $pClick_{device}$) of RR, scaled SNM and clipped SNM compared to the greedy non-private case and the greedy non-personalized baselines. The left column shows auctions with relatively few candidates (3). Here, we see that RR is the clear winner at all values of $\epsilon$. However, the right column shows auctions with more candidates (10 or more). Here, clipped SNM can outperform RR at very low $\epsilon$, but it barely improves on the non-personalized baseline and can be catastrophically bad if the public scores are less informative (2nd row). In the 3rd row, we show examples where using the public scores already suffices to choose the best candidate; for 3 candidates, RR is best for all $\epsilon$, and for 10 candidates, clipped SNM is best for all $\epsilon$.

\begin{figure}[ht]
\vskip 0.2in
\begin{center}
\centerline{\includegraphics[width=\columnwidth]{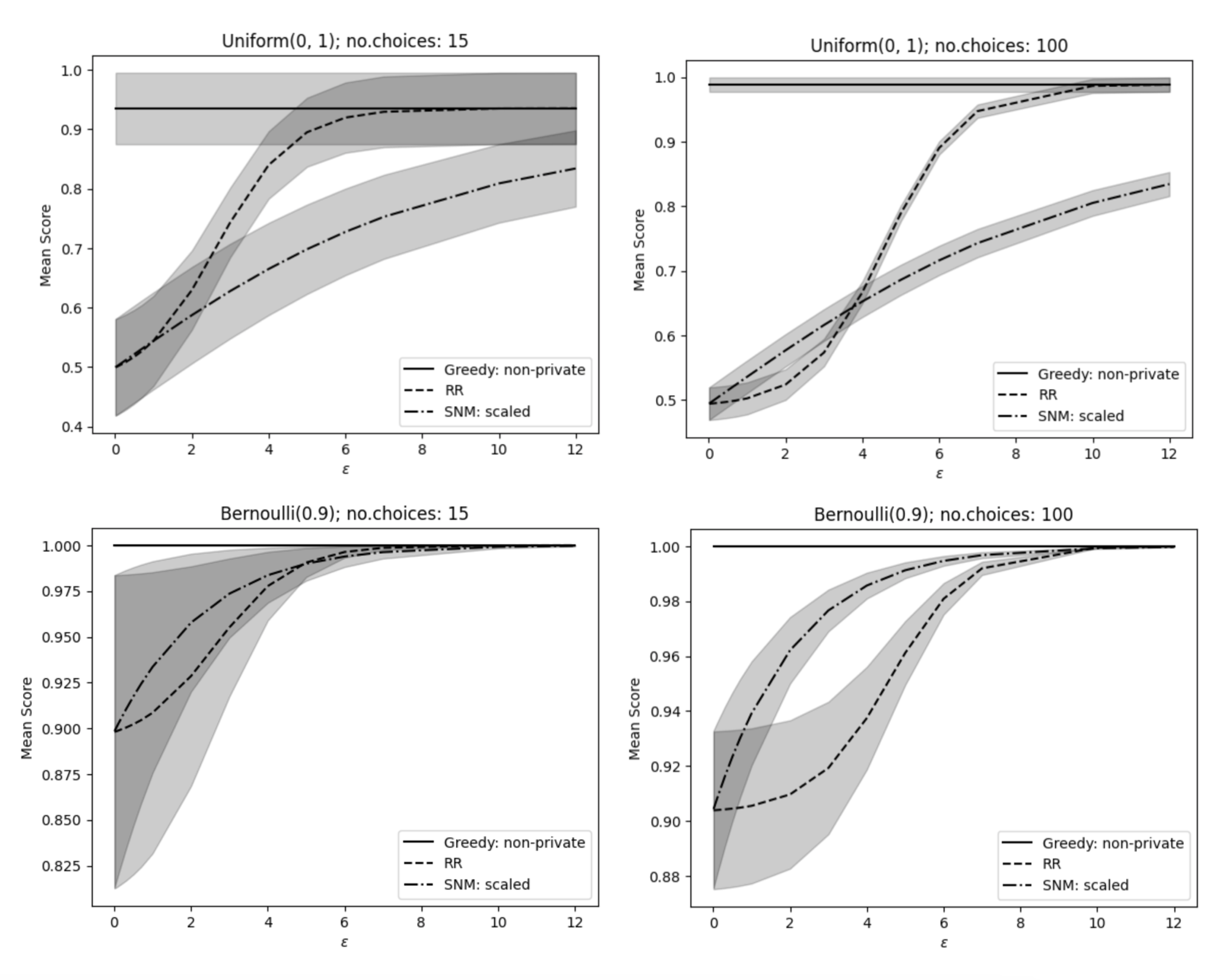}}
\caption{Comparison of the performance of RR vs scaled SNM for synthetic data. Synthetic scores are generated 100 times, and shaded region shows one standard deviation in realized score. a) 15 candidate scores are sampled from a uniform distribution, b) 100 candidate scores are sampled from a uniform distribution, c) 15 candidates scores are sampled from a binomial distribution with $p=0.9$, d) 100 candidate scores are sampled from a binomial distribution with $p=0.9$ . }
\label{synthetic-scores}
\end{center}
\vskip -0.2in
\end{figure}

Next, in Figure \ref{synthetic-scores}, we compare RR and SNM's performance on synthetically generated scores. From our previous analysis, we hypothesize that both the score distribution as well as the number of choices available will affect which algorithm is most performant. For this analysis, we ignore the possibility of informative public scores and use scaled SNM. We sample 100 sets of scores from the four specified distributions and plot the mean expected value of each method and shade +/- one standard deviation. We observe that as candidate set sizes (number of choices) increase, the relative performance of randomized response degrades, as we would expect given the functional dependence of RR but not SNM on $a$. We do also see effects of different score distributions---at smaller candidate sizes, RR outperforms scaled SNM when the scores are sampled from a uniform distribution, but SNM outperforms RR at small epsilons when scores are sampled from a weighted Bernoulli (a more extreme distribution).

\section{Comparison of results on the Taobao and internal dataset}
\label{taobao-vs-internal-results}
We hypothesize that the more pronounced dominance of RR in the Alibaba dataset occurs due to two key differences between the datasets. First, the distributions of candidates per auction in the two datasets are quite different. In the Alibaba dataset, on average, there are only 4 candidates per auction, which is significantly less than the internal dataset. We know that RR will have higher expected utility as the number of candidates decreases (Equation \ref{equation-rr}), so it makes sense that RR would generally be more dominant in the Alibaba dataset.

\begin{figure*}[ht]
\vskip 0.2in
\begin{center}
\centerline{\includegraphics[width=\textwidth]{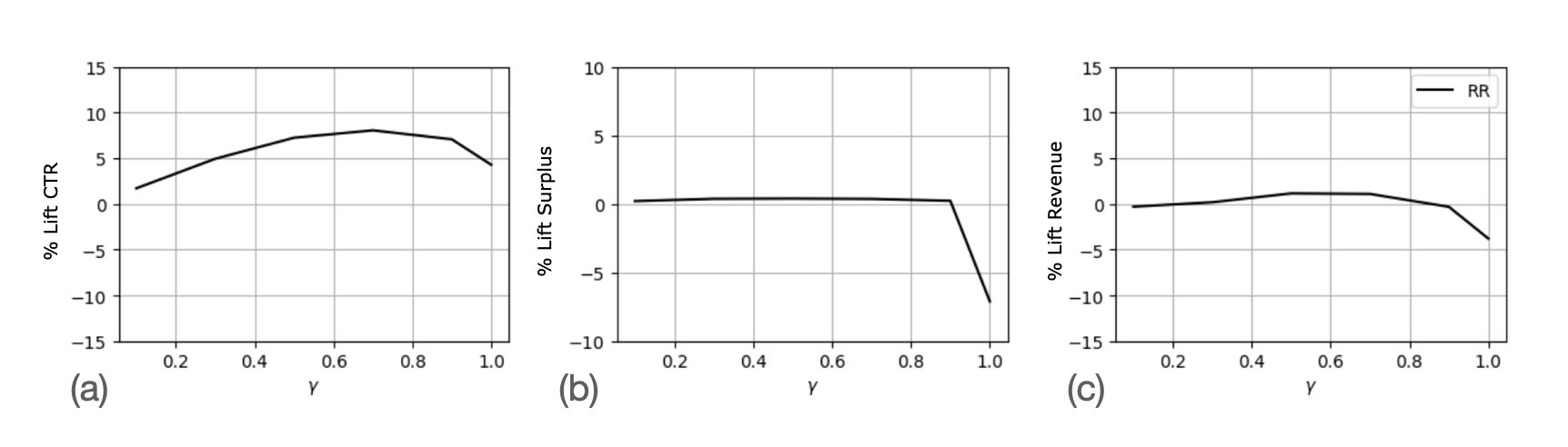}}
\caption{Effects of the changing the cutoff parameter, $\gamma$ on (a) CTR, (b) surplus, and (c) revenue on the internal datset.}
\label{internal-rr-cutoff}
\end{center}
\vskip -0.2in
\end{figure*}

Second, we would expect that in search (internal dataset) vs display (Alibaba dataset) advertising, the information gain from $pClick_{server}$ to $pClick_{device}$ should be relatively smaller; in search ads, $pClick_{server}$ is already informed by the search term, which is a strong signal of user intent. This hypothesis also explains the difference we see in the behavior in Figure \ref{rr-cutoff} compared to our internal data (Figure \ref{internal-rr-cutoff}): the process appears to be much more limited by recall in the Alibaba data, meaning it is generally better to send more items to the device. Indeed, we find that, if we consider the distribution of $|pClick_{device} - pClick_{public}|$ for our internal dataset compared to the Alibaba dataset, the median of the Alibaba dataset is more than 100\% higher than that of the internal dataset. 

\begin{figure}[ht]
\vskip 0.2in
\begin{center}
\centerline{\includegraphics[width=0.7\columnwidth]{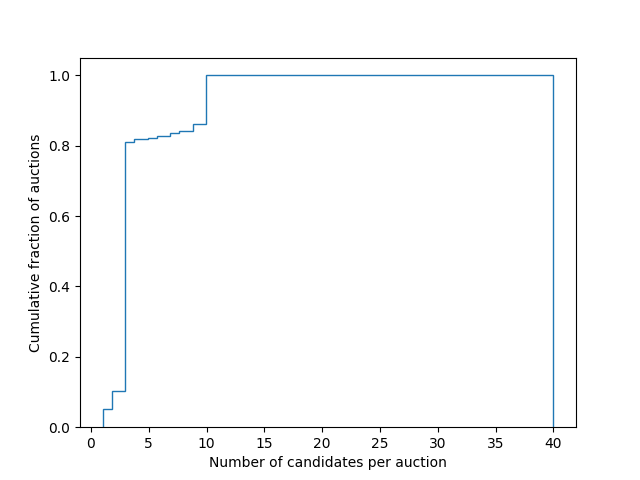}}
\caption{Cumulative distribution function of the number of candidates per auction in the public dataset.}
\label{distribution}
\end{center}
\vskip -0.2in
\end{figure}

Given these observations, we modify the Alibaba dataset to be more similar to the internal dataset and observe the changes in CTR trends for RR vs SNM. To modify the number of candidates per auction, we split the dataset into three groups based on the cumulative distribution function of candidate frequencies (Figure \ref{distribution}): auctions with three or fewer candidates, between three and ten candidates, and ten or more candidates.  Next, to modify the information gain between $pClick_{server}$ and $pClick_{device}$, we mix varying fractions of $pClick_{server}$ into $pClick_{device}$ to achieve a new $pClick_{device}$ that is more similar to the server estimate:

\begin{equation}
\label{pclick-mixing}
    pClick_{device} = \alpha \times pClick_{device} + (1-\alpha) \times pClick_{server}.
\end{equation}

\begin{figure}[ht]
\vskip 0.2in
\begin{center}
\centerline{\includegraphics[width=\columnwidth]{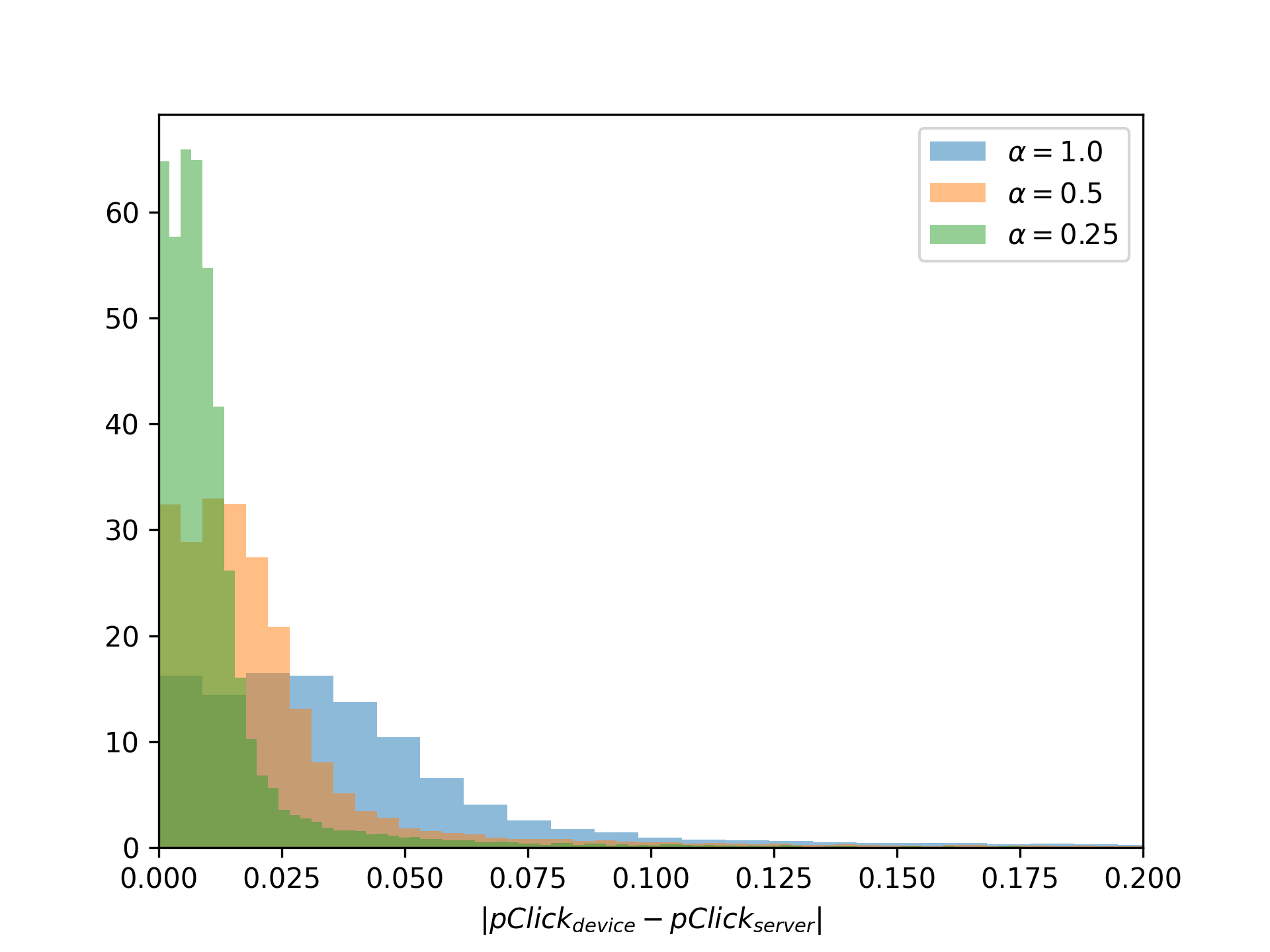}}
\caption{The distribution of $|pClick_{device}-pClick_{server}|$ resulting from mixing in server pClick scores as in Equation \ref{pclick-mixing}.}
\label{delta_pclick}
\end{center}
\vskip -0.2in
\end{figure}

\begin{figure}[ht]
\vskip 0.2in
\begin{center}
\centerline{\includegraphics[width=\columnwidth]{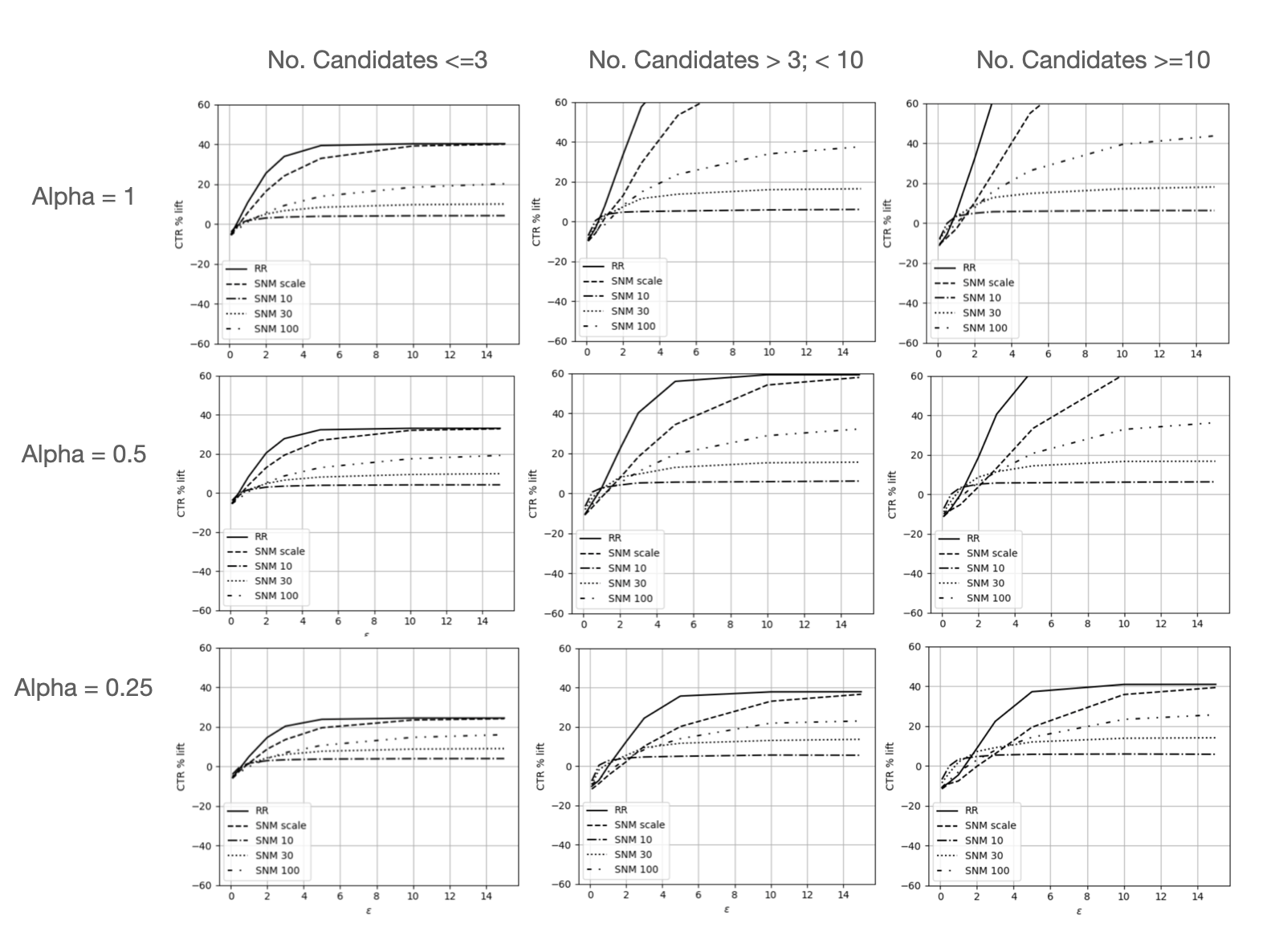}}
\caption{Effects on CTR of varying the number of candidates per auction (columns) and the similarity of server and device pClick scores (rows). Decreasing $\alpha$ means more similarity.}
\label{alpha_no-candidates}
\end{center}
\vskip -0.2in
\end{figure}

We show the resulting $|pClick_{device}-pClick_{server}|$ histograms in Figure \ref{delta_pclick}, and the results of using these modified $pClick_{device}$ values in the ads auctions are shown in Figure \ref{alpha_no-candidates}. We observe that, both as the number of candidates per auction grows (columns) and as $pClick_{device}$ gets closer to $pClick_{server}$ (rows), SNM with clipping becomes more dominant at small $\epsilon$, similar to the internal dataset.

\begin{figure}[ht]
\vskip 0.2in
\begin{center}
\centerline{\includegraphics[width=\columnwidth]{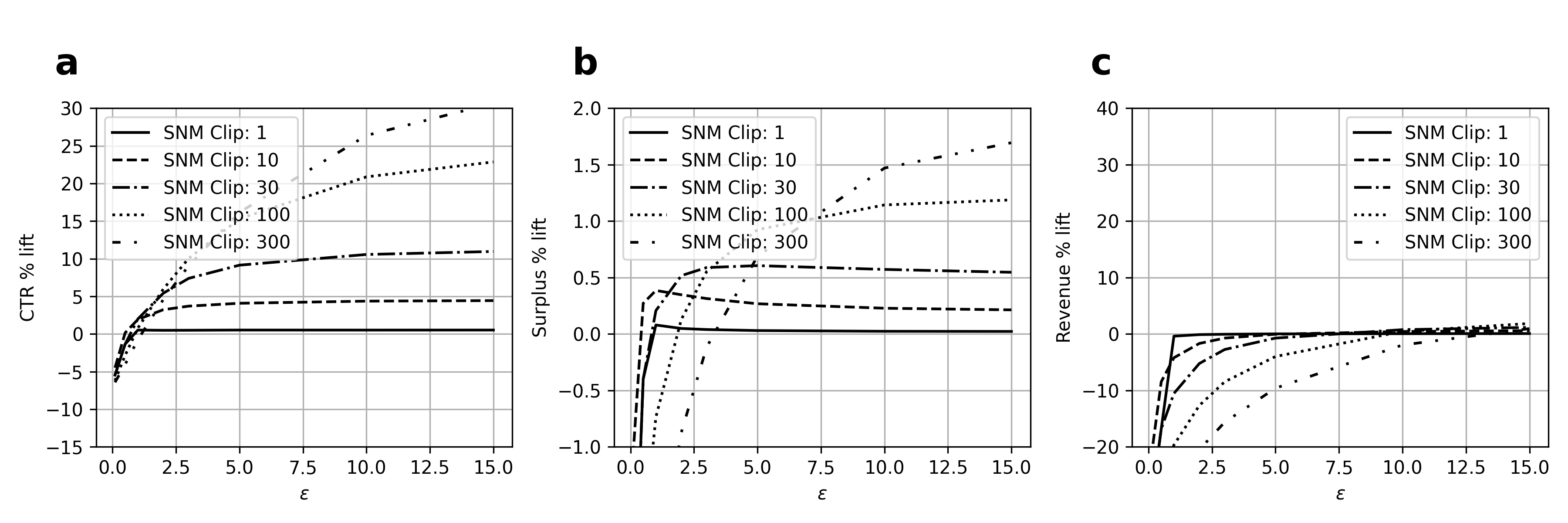}}
\caption{Effects of changing the clipping bound in the SNM mechanism on the Taobao dataset.}
\label{taobao-clipping-bound}
\end{center}
\vskip -0.2in
\end{figure}

\begin{figure*}[ht]
\vskip 0.2in
\begin{center}
\centerline{\includegraphics[width=\textwidth]{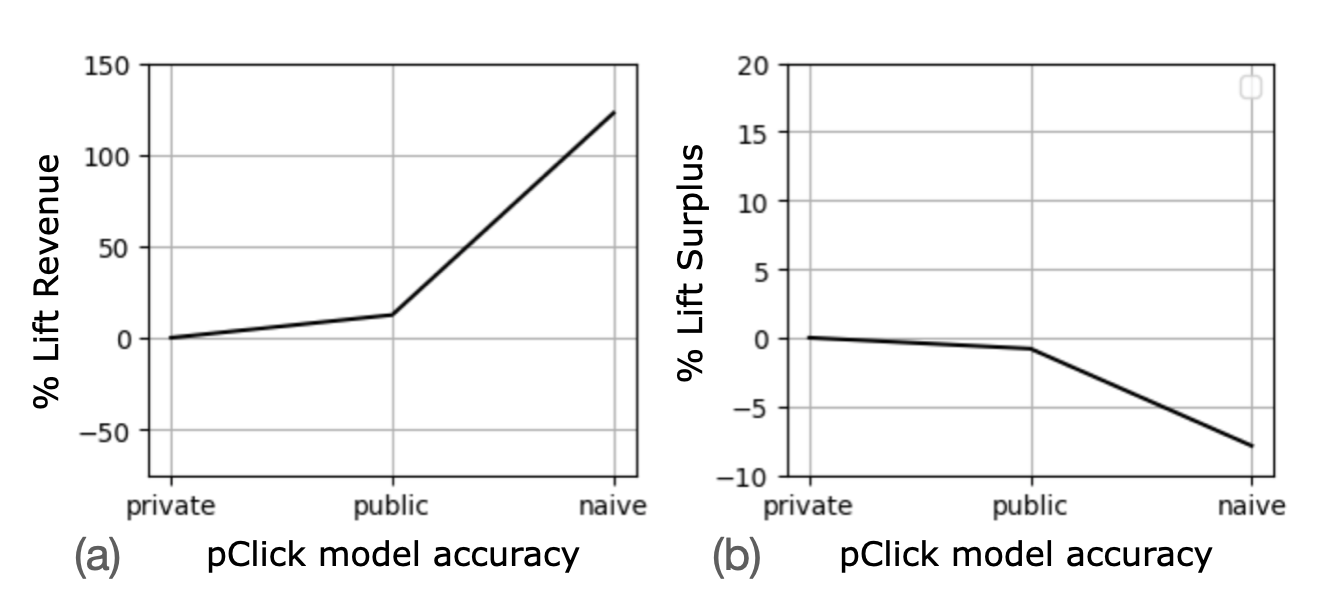}}
\caption{Effects of the changing the $pClick$ model accuracy on platform revenue and advertiser surplus on the internal datset.}
\label{internal-pricing}
\end{center}
\vskip -0.2in
\end{figure*}

\begin{figure*}[ht]
\vskip 0.2in
\begin{center}
\centerline{\includegraphics[width=\textwidth]{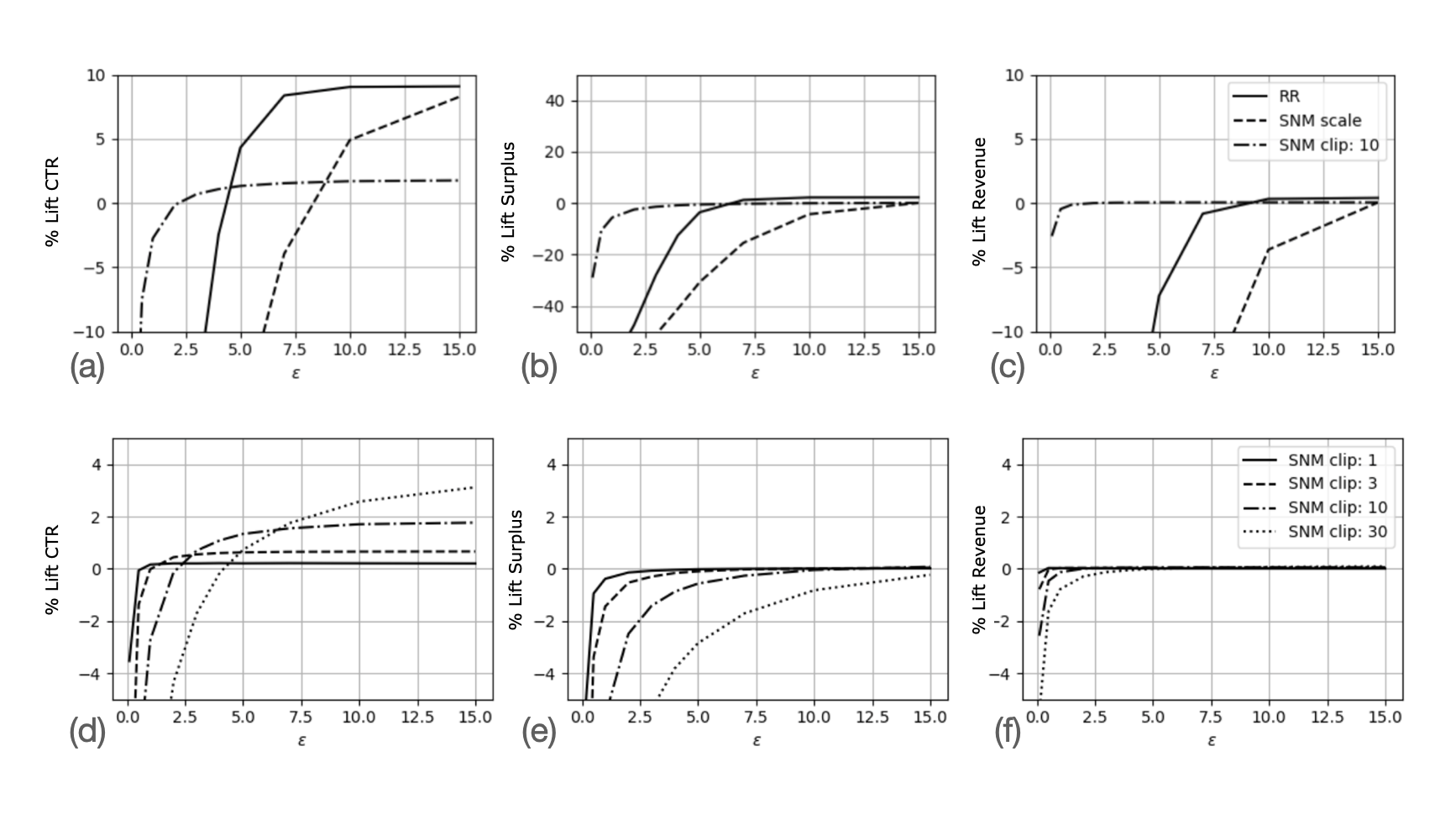}}
\caption{Effects of the privacy parameter, $\epsilon$ on (a) CTR, (b) surplus and (c) revenue for RR and SNM with clipping bound = 10 on the internal datset. (d)-(f) show the effects of changing the clipping bound in SNM.}
\label{internal-rnm-vs-rr}
\end{center}
\vskip -0.2in
\end{figure*}

\end{document}